\documentclass[sigconf,screen]{acmart}

\AtBeginDocument{%
  }

\setcopyright{cc}
\setcctype{by}
\acmDOI{10.1145/3832783.3834421}
\acmYear{2026}
\copyrightyear{2026}
\acmISBN{979-8-4007-2882-2/2026/10}
\acmConference[ASE '26]{Proceedings of the 41st IEEE/ACM International Conference on Automated Software Engineering}{October 12--16, 2026}{Munich, Germany}
\acmBooktitle{Proceedings of the 41st IEEE/ACM International Conference on Automated Software Engineering (ASE '26), October 12--16, 2026, Munich, Germany}
\acmSubmissionID{ase26main-p2670-p}
\received{2026-03-26}
\received[accepted]{2026-06-18}

\usepackage{microtype}
\usepackage{subcaption}
\usepackage{booktabs}
\usepackage{graphicx}
\usepackage{tcolorbox}
\newtcolorbox{findingbox}[1][]{
    colframe=gray,
    colback=gray!5,
    boxrule=1pt,
    arc=5pt,
    left=5pt,
    right=5pt,
    top=5pt,
    bottom=5pt,
}
\tcbuselibrary{listings}
\usepackage{multirow}
\usepackage{fvextra}
\fvinlineset{
    breaklines=true,
    breakanywhere=true,
    breaksymbolleft={},
    breaksymbolright={},
    breakanywheresymbolpre={}
}

\usepackage{etoolbox}
\newcounter{bibitemcount}
\AtBeginEnvironment{thebibliography}{%
  \setcounter{bibitemcount}{0}%
  \let\dcawareorigbibitem\bibitem
  \RenewDocumentCommand{\bibitem}{o m}{%
    \stepcounter{bibitemcount}%
    \ifnum\value{bibitemcount}=54\relax
      \typeout{DCAware: forcing column break before bibliography item 54}%
      \newpage
    \fi
    \IfNoValueTF{#1}
      {\dcawareorigbibitem{#2}}
      {\dcawareorigbibitem[#1]{#2}}%
  }%
}

\begin{document}

\title{Escaping the Self-Repair Trap: Improving Test Oracle Generation via Dual-Context Awareness}


\author{Kefan Li}
\email{kefanli@buaa.edu.cn}
\orcid{0009-0000-1824-8889}
\affiliation{%
  \institution{Beihang University}
  \department{School of Computer Science and Engineering}
  \country{China}
}

\author{Hongyue Yu}
\email{Natt1e@buaa.edu.cn}
\orcid{0009-0002-4569-941X}
\affiliation{%
  \institution{Beihang University}
  \department{National College for Excellent Engineers}
  \country{China}
}

\author{Yuan Yuan}
\authornote{Corresponding author.}
\email{yuan21@buaa.edu.cn}
\orcid{0000-0003-4233-4407}
\affiliation{%
  \institution{Beihang University}
  \department{School of Computer Science and Engineering and Qingdao Research Institute and Hangzhou Innovation Institute}
  \country{China}
}

\begin{abstract}
Large Language Models (LLMs) have shown strong potential for regression-oracle completion, where a test prefix is given and the current program version is treated as expected behavior. 
Recent approaches increasingly rely on iterative self-repair and execution feedback, but optimizing execution success does not necessarily yield strong fault-revealing oracles. 
This objective, widely adopted in repair-based methods, serves only as a proxy and may be misaligned with the true goal of oracle generation. 
Such misalignment biases the repair process, giving rise to a feedback-driven degeneration that we term the \emph{Self-Repair Trap}, where iterative repair progressively drives models toward assertions that are easier to satisfy but less effective at detecting faults. 
To address this issue, we propose \textbf{DCAware}, a computationally efficient, non-iterative framework that prioritizes high signal-to-noise contextual grounding over multi-round repair.
DCAware integrates structured static context with selectively 
retrieved dynamic states, enabling precise and robust oracle generation without iterative feedback loops. 
Extensive experiments based on execution and mutation testing show that DCAware consistently improves fault-revealing effectiveness while maintaining high execution success, outperforming prior methods with substantially lower computational cost. 
Our results suggest that improving contextual quality is more effective than adding iterative repair complexity in the studied regression-oracle setting. 
\end{abstract}

\ccsdesc[500]{Software and its engineering~Software testing and debugging}

\keywords{Oracle Generation, Software Testing, Large Language Models}

\maketitle

\section{Introduction}

Unit tests play a crucial role in ensuring the reliability, maintainability, and evolvability of software systems \cite{godefroid2005dart,runeson2006survey}. 
They help developers detect regressions early, preserve behavioral consistency across code revisions, and act as executable documentation that facilitates program comprehension, refactoring, and long-term maintenance \cite{meszaros2007xunit}. 
In modern software engineering practice, the availability and quality of unit tests are widely regarded as key indicators of project health and development maturity.
However, writing high-quality unit tests is notoriously labor-intensive and time-consuming \cite{daka2014survey}. 
A typical unit test consists of two essential components: the \emph{test prefix}, which prepares the environment and invokes the target code, and the \emph{test oracle}, which determines the expected behavior or correctness condition \cite{barr2015oracle}. 
Compared with generating test prefixes, producing accurate and meaningful test oracles is more challenging due to the need for semantic understanding, behavioral reasoning, and inference about program intent. 
In this work, we focus on \emph{regression-oracle generation} in a test-completion setting, where an executable test prefix is already available and the current program version is assumed to represent the expected behavior. 
The goal is to automatically complete or strengthen the missing oracle assertion rather than generate complete test cases from scratch.

Therefore, regression-oracle generation remains a frequently studied and highly challenging problem.
Early methods \cite{atlas,mastropaolo2021studying} attempted to learn assertion patterns directly from code but were significantly constrained. 
Subsequently, transformer-based methods emerged, following two distinct trajectories: rule-based generation with model-based reranking (e.g., TOGA \cite{toga}), and direct code generation via fine-tuned small-scale models (e.g., TECO \cite{teco}, LLM-AG \cite{llm-ag}, TOGLL \cite{togll}). 
Unfortunately, these small models are fundamentally restricted in reasoning, and many were trained on EvoSuite-generated data \cite{evosuite}, which differs significantly from human-written code conventions, limiting their real-world utility.

Recently, Large Language Models (LLMs) have shown impressive generalization across software domains. 
Inspired by these advancements, recent literature has explored incorporating LLMs into oracle generation \cite{chatassert,llm-ag,actual_vs_expected,molinelli2025llms}. 
Notably, ChatAssert \cite{chatassert} introduced a paradigm that bypasses small model training, utilizing prompts, vector retrieval, and iterative execution feedback in a self-repair workflow to achieve high accuracy. 
By employing modern LLMs, ChatAssert significantly outperformed previous approaches.

\begin{figure}[ht]
\centering
\includegraphics[width=0.99\linewidth]{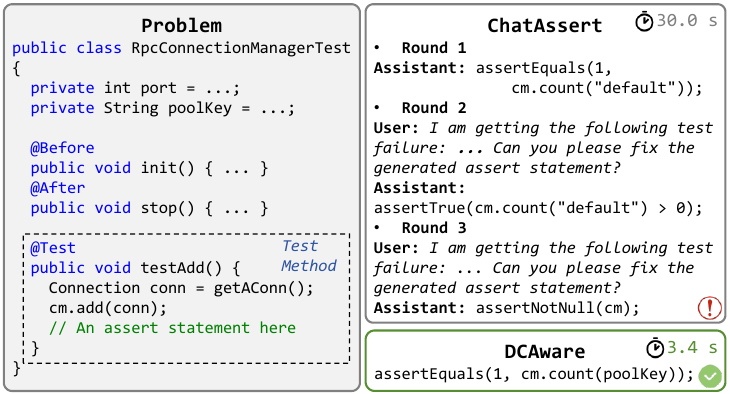}
\caption{A motivating example illustrating the degeneration of test oracle quality during iterative repair.}
\Description{A motivating example illustrating the degeneration of test oracle quality during iterative repair.}
\label{fig:motivation}
\end{figure}

Despite these advancements, a potential limitation lies in the objective underlying execution-guided repair approaches. 
Existing methods typically optimize for execution success, ensuring that generated assertions compile and pass runtime checks. 
However, execution success is only a proxy and may not always align with the goal of generating strong fault-revealing assertions. 
This objective misalignment can bias the repair process. 
When guided by execution feedback, models may modify failing assertions toward easier-to-satisfy conditions rather than improving their fault-revealing capability. 
As a result, iterative repair can lead to a feedback-driven degeneration, where some assertions become easier to satisfy but less effective at revealing faults. 
We term this phenomenon the \emph{Self-Repair Trap}. 
As illustrated in Figure~\ref{fig:motivation}, although repaired assertions achieve execution success, their fault-detection capability can be significantly weakened.
In addition, these approaches often incur substantial computational overhead due to iterative interactions and repeated executions.

To address this limitation, we hypothesize that high-quality test
oracle generation does not require multi-round trial-and-error reasoning. 
Instead, it depends on providing high signal-to-noise contextual grounding in a non-iterative pipeline.
We argue that by supplying carefully denoised static structures and precisely targeted dynamic states, the model can directly generate fault-revealing assertions without relying on iterative repair.

Inspired by recent agentless software-engineering workflows that reduce unnecessary multi-round agent interactions through structured context construction \cite{xia2024agentlessdemystifyingllmbasedsoftware, li2025patchpilotcostefficientsoftwareengineering}, 
we propose \textbf{DCAware}, 
a computationally efficient dual-context approach for regression-oracle generation. 
DCAware integrates two key components: \emph{Contextual Semantic Folding}, which constructs a concise structural representation by filtering irrelevant code, and \emph{Intent-Driven Dynamic State Extraction}, which selectively retrieves runtime states essential for oracle reasoning. 
Together, these components provide high-quality contextual grounding that enables precise oracle generation in a non-iterative pipeline.
We evaluate DCAware using execution-based metrics and mutation testing via \verb|PITest| \cite{pitest}. 
Results show that DCAware improves fault-revealing effectiveness while maintaining high execution success, and achieves these gains with substantially lower computational cost than prior methods.

In summary, the main contributions of this paper are as follows:
\begin{enumerate}
\item We identify an objective misalignment in execution-guided oracle generation and characterize its effect as the \emph{Self-Repair Trap}, a feedback-driven degeneration that weakens fault-revealing capability.
\item We propose \textbf{DCAware}, a computationally efficient, non-iterative approach that improves oracle generation through high-quality static and dynamic contextual grounding.
\item We conduct extensive dynamic evaluations showing that DCAware achieves superior fault-revealing effectiveness with substantially lower computational cost than state-of-the-art methods.
\end{enumerate}

\section{Methodology}
\begin{figure*}[h!]
\includegraphics[width=0.99\linewidth]{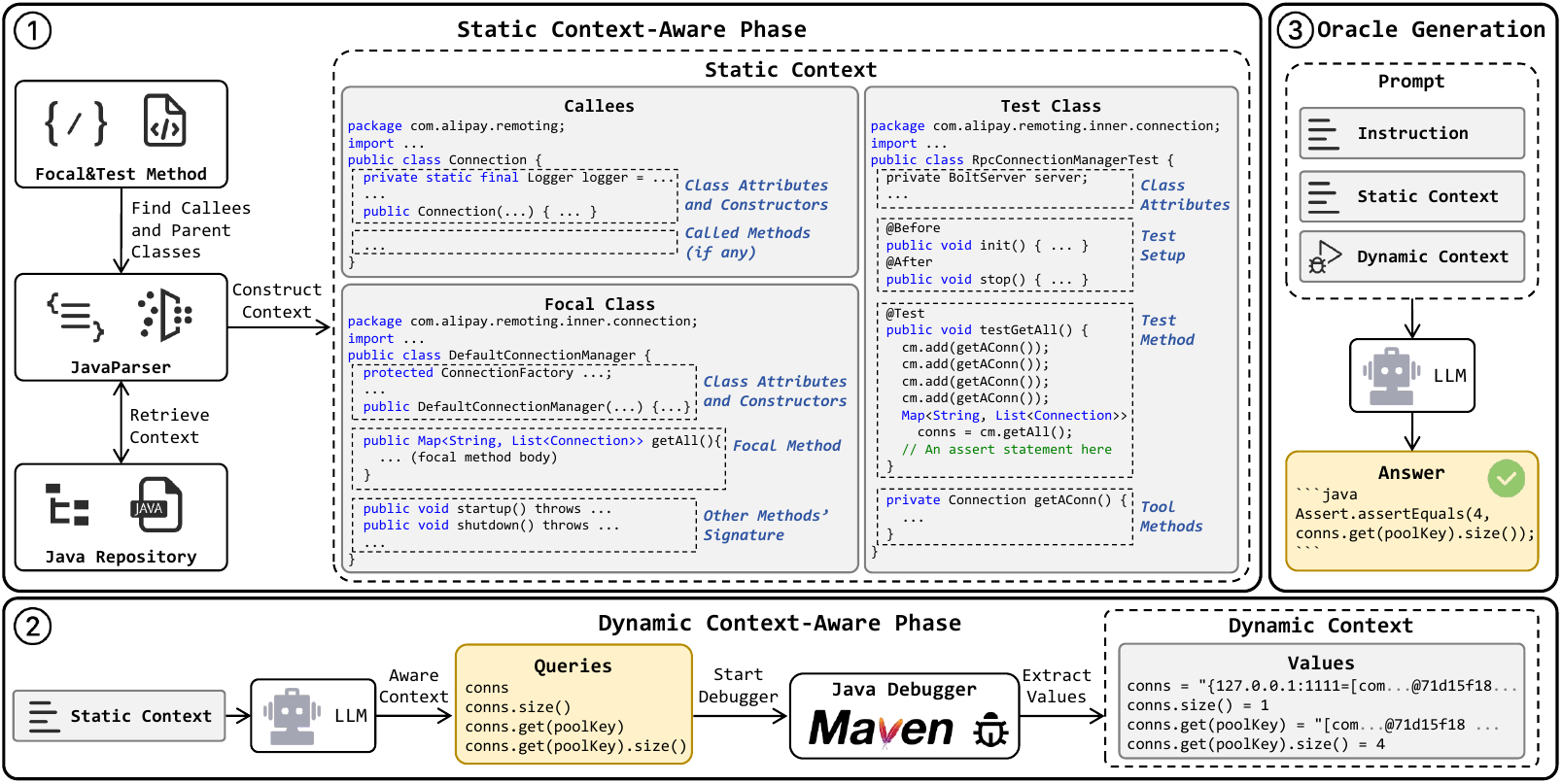}
\caption{Overview of the DCAware framework. The pipeline constructs a folded static context, retrieves targeted runtime states through LLM-generated debugger queries, and combines both contexts to generate the oracle in a single pass.}
\Description{Overview of the DCAware framework. The pipeline constructs a folded static context, retrieves targeted runtime states through LLM-generated debugger queries, and combines both contexts to generate the oracle in a single pass.}
\label{fig:overview}
\end{figure*}

In this section, we detail the architectural design of the DCAware approach. 
Inspired by the philosophy of information denoising, our pipeline avoids complex, multi-round agent loops. 
Instead, it is structured into three distinct and sequential phases: Static Context-Aware, Dynamic Context-Aware, and Oracle Generation. 
The overall procedure is illustrated in Figure~\ref{fig:overview}.

\subsection{Static Context-Aware Phase}

The Static Context-Aware phase is designed to gather sufficient structural code information while actively controlling token noise. 
Previous approaches typically provided only the focal method, test setup, and test prefix. 
This restrictive view omits critical contextual elements necessary for semantic understanding.

To resolve this informational deficit without overwhelming the LLM's attention mechanism, we introduce \textbf{Contextual Semantic Folding}. 
LLMs frequently suffer from the ``lost in the middle'' phenomenon when processing long code contexts; massive amounts of irrelevant local variables and uninvoked method bodies act as noise that dilutes the model's focus. 
By utilizing \verb|JavaParser| \cite{javaparser}, we extract three specific dimensions of static context and apply precise semantic folding:

\textbf{Callees:} We extract all methods invoked within the test prefix. For each identified class, we retain import statements and the structural skeleton. We preserve complete method bodies only for explicitly invoked methods; all other method bodies are strictly deleted (folded).
\textbf{Focal Class:} We preserve the import statements and the full implementation of the target focal method. Crucially, we retain the signatures of all other sibling methods within the focal class while folding their bodies. This provides the LLM with a comprehensive view of the class's capabilities (its skeleton) without introducing the noise of unrelated implementations.
\textbf{Test Class:} The golden oracle is replaced with a placeholder comment. To prevent data leakage and further compress the prompt, we completely fold all other test methods. We retain the complete bodies of utility and lifecycle configuration methods (e.g., \verb|@Before|), as these initialization mechanisms are essential for understanding the testing environment.

This Contextual Semantic Folding effectively improves the context's Signal-to-Noise Ratio (SNR), grounding the LLM in a solid structural reality.
Furthermore, this structural reality serves as an indispensable blueprint for the LLM to formulate precise dynamic queries in the subsequent phase.

\subsection{Dynamic Context-Aware Phase}

Understanding the runtime state is frequently necessary for asserting complex object behaviors. 
To achieve this, we introduce \textbf{Intent-Driven Dynamic State Extraction}. 
Unlike existing methods that passively dump all available local variables into the prompt—a practice that exacerbates context noise—we treat the LLM as an \emph{Active Debugger}. 

Initially, the folded static context is provided to the LLM. 
The model analyzes the code skeleton to formulate an intent, outputting a specific list of queries (variables, mathematical expressions, or side-effect-free function calls) whose runtime values are required to write a precise oracle. 
Once the LLM generates this query list, we execute an automated debugging sequence. 
We inject a valid suspension point (\verb|int __breakpoint__ = 0;|), spawn a Maven test process configured for remote JVM debugging, and attach the Java Debugger (JDB). 
Through automated pseudo-terminal interaction, we command the JDB to evaluate the exact queries formulated by the LLM. 
If a query fails due to invalid syntax, out-of-scope variables, or runtime exceptions, JDB returns the corresponding error message, which is recorded and provided as feedback while the remaining queries continue to execute. 
The successfully retrieved values provide the dynamic context, ensuring the LLM only receives high-entropy, semantically relevant runtime information. 

\subsection{Oracle Generation}
In the final phase, the folded static context and the intent-driven dynamic context are synthesized.
Through explicit instructions, we direct the model to generate an accurate test oracle in a non-iterative manner.
To prevent the model from outputting overfitted or fragile oracles, we establish \textbf{Anti-Overfitting Guardrails} within the prompt. 
During dynamic execution, the runtime environment might return volatile memory references (e.g., \verb|obj@12345|) or transient timestamps. Without constraints, LLMs tend to generate fragile, exact-match assertions that pass the immediate execution check but inevitably fail in subsequent regression tests. 

\begin{figure}[h!]
\centering
\includegraphics[width=0.99\linewidth]{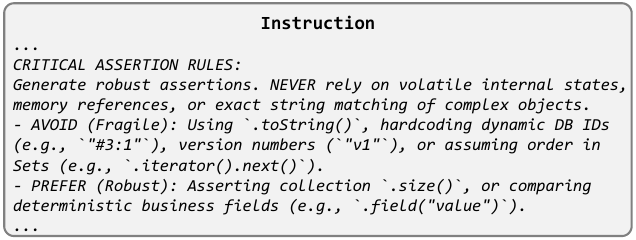}
\caption{Anti-overfitting guardrails used during oracle generation.}
\Description{Anti-overfitting guardrails used during oracle generation.}
\label{fig:inst}
\end{figure}

Therefore, these guardrails strictly prohibit matching against dynamically allocated addresses, forcing the model to extract and assert semantically stable attributes (e.g., utilizing \verb|.size()| or deterministic business fields). 
The specific instruction design is illustrated in Figure~\ref{fig:inst}.

The LLM outputs the final test oracle encapsulated within a Markdown Java code block, entirely bypassing the need for a secondary execution and repair loop.

\section{Results}
In this section, we present the experimental setup and answer the formulated research questions.

\subsection{Research Questions}
We propose the following research questions:
\begin{itemize}
\item \textbf{RQ1:} How effective are the oracles generated by DCAware?
\item \textbf{RQ2:} How does each static-context component contribute to the final performance of DCAware?
\item \textbf{RQ3:} How does each dynamic-context strategy affect the final performance of DCAware?
\item \textbf{RQ4:} How does DCAware perform in terms of token consumption and execution time?
\item \textbf{RQ5:} How do the static and dynamic contexts complement each other in the complete DCAware pipeline?
\end{itemize}

\subsection{Experimental Setting}
\textbf{Datasets:} Due to the substantial computational cost of dynamic evaluation, we utilize a subset of 500 distinct problems from the TECO dataset, exactly identical to the official ChatAssert evaluation subset \cite{chatassert}. 
Here, each problem represents a test-method assertion-completion instance rather than a project. 
These instances are collected from 38 open-source Java projects and provide executable test prefixes with missing oracle assertions. 
We verified that all golden (human-written) oracles successfully execute and pass on these problems, ensuring baseline validity. 
For consistency with TECO and ChatAssert, we follow their single-assertion completion setting, where each instance requires generating one oracle assertion. Handling multiple assertions within a single test case is beyond the scope of this work. 

\textbf{Baselines:} We compare DCAware against seven baselines across three categories:
\begin{itemize}
\item \textbf{Training-based:} \textbf{TECO} \cite{teco} (CodeT5 \cite{codet5}), \textbf{TOGA} \cite{toga} (CodeBERT \cite{codebert}), \textbf{LLM-AG} \cite{llm-ag} (CodeT5 \cite{codet5}), and the top two performing variants (CodeGen \cite{codegen}, CodeParrot \cite{codeparrot}) of \textbf{TOGLL} \cite{togll}. 
We exclude Doc2OracLL \cite{hossain2025doc2oracll} as its source code and models are unavailable, though its methodology resembles TOGLL.
\item \textbf{LLM-based:}
\textbf{ChatAssert} \cite{chatassert} and a single-query \textbf{Direct Prompting} baseline. 
\item \textbf{Rule-based:} \textbf{Trivial} generates solely \verb|assertTrue(true);|, serving as a fundamental lower bound for oracle strength to demonstrate the necessity of semantic validation.
\end{itemize}
Notably, while prior works (e.g., TECO and ChatAssert) originally replaced string literals with a generic \verb|STR| token to reduce generation difficulty, we strictly retain the original strings across all baselines to maintain realistic testing conditions.

\textbf{Models:} We employ \textit{Qwen3-Coder-30B-A3B-Instruct-FP8} \cite{qwen3technicalreport} (deployed via vLLM \cite{vllm} on a single 40\,GB NVIDIA A100 GPU) and \textit{GPT-5-mini} \cite{openai_gpt5} (accessed via Azure API). 
In the following experiments, we refer to these models as \textit{Qwen3-Coder-30B} and \textit{GPT-5-mini}, respectively.
To prevent data contamination and latency inaccuracies from shared prompt caching, each evaluated approach uses a unique \verb|cache_salt| to strictly isolate cache memory.

\textbf{Metrics:} Prior works often rely on static metrics (e.g., Exact Match, ROUGE \cite{rouge}, CodeBLEU \cite{codebleu}), which fail to accurately capture functional correctness. 
To address this, we employ fully dynamic execution-based metrics:

\begin{itemize}
\item \textbf{Pass Rate (\%):} The percentage of generated oracles that the Maven test process completes normally without runtime crashes, and the extracted test results confirm that the target test method successfully passes.
\item \textbf{Kill Golden (KG) (\%):} Measures fault-revealing capability using PITest \cite{pitest} (with the \verb|DEFAULTS| mutator). 
Since high Pass Rates can be artificially inflated by trivial oracles, mutation analysis provides rigorous semantic validation. 
We strictly filter the dataset to exclude problems where PITest fails, or the golden oracle kills zero mutants, yielding $N=362$ valid problems. 
KG is calculated as the macro-average of individual kill rates:
\begin{equation}
KG = \frac{1}{N} \sum_{i=1}^{N} \frac{|M_{killed,i}^{gen} \cap (M_{killed,i}^{golden} \cup M_{survived,i}^{golden})|}{|M_{killed,i}^{golden}|}
\end{equation}
where $M_{killed,i}^{gen}$ represents mutants killed by the generated oracle (scoring zero if execution fails). 
KG evaluates generated-oracle effectiveness only within the behavioral scope exercised by the golden oracle, defined by the union of golden-killed and golden-survived mutants. 
This design avoids inflating scores with mutants outside the tested behavior scope.

\item \textbf{Shared Kill Golden (Shared KG) (\%):}
We compute Shared KG on the mutation-valid instances passed by all methods in each comparison, eliminating bias from different Pass Rates. 
The \textbf{Trivial} baseline serves as a lower bound because it can score non-zero when mutants are killed by the test prefix before the assertion is reached; the improvement over this baseline reflects the assertion's additional fault-revealing contribution. 
Because the shared set is recomputed for each table, Shared KG values are comparable only within the same table. 
\item \textbf{Runtime Error (RE) (\%):} The ratio of generated oracles that crash during execution (e.g., syntax errors, unresolved dependencies, or exceptions). 
\item \textbf{Test Failed (TF) (\%):} The proportion of generated oracles where the Maven test process completes normally without any runtime crashes, but the extracted test results indicate that the target test method failed. 
\item \textbf{Empty Oracle (EO) (\%):} The percentage of cases where the method fails to output any oracles.
\end{itemize}

\textbf{Execution Environment:} All evaluations are conducted within isolated Docker containers to ensure stable and reproducible dynamic assessments.

\subsection{RQ1: Performance of DCAware}
\label{subsection:rq1}

\begin{table*}[ht!]
\caption{Execution-based metrics comparison of DCAware and other baselines. The best-performing results are in bold.}
\label{tab:rq-main}
\setlength{\tabcolsep}{12pt}
\begin{tabular}{llccccc}
\toprule
Method & Model & Pass Rate (\%) & KG (\%) & RE (\%) & TF (\%) & EO (\%) \\
\midrule
TOGA \cite{toga}        & CodeBERT & 23.00 & 14.54 & 1.00  & 12.60 & 63.40 \\
TOGLL \cite{togll}      & CodeGen-350M & 26.20 & 18.37 & 35.60 & 24.00 & 14.20 \\
TOGLL \cite{togll}      & CodeParrot-110M & 22.60 & 16.13 & 40.80 & 23.80 & 12.80 \\
LLM-AG \cite{llm-ag}    & CodeT5 & 8.80 & 8.75 & 73.80 & 16.40 & 1.00 \\
TECO \cite{teco}        & CodeT5 & 25.60 & 24.19 & 24.00 & 21.20 & 29.20 \\
\midrule
Direct Prompting 
& Qwen3-Coder-30B   & 55.00 & 50.76 & 21.60 & 23.40 & 0.00 \\
& GPT-5-mini        & 61.40 & 54.58 & 21.40 & 17.20 & 0.00 \\
ChatAssert \cite{chatassert}
& Qwen3-Coder-30B   & 68.40 & 59.00 & 28.40 & 0.80 & 2.40 \\
& GPT-5-mini        & 73.20 & 61.56 & 25.40 & 0.20 & 1.20 \\
DCAware
& Qwen3-Coder-30B   & 80.80 & 73.20 & 7.40 & 11.80 & 0.00 \\
& GPT-5-mini        & \textbf{88.20} & \textbf{78.17} & 4.20 & 7.60  & 0.00 \\
\bottomrule
\end{tabular}
\end{table*}

As shown in Table~\ref{tab:rq-main}, methods trained primarily on EvoSuite-generated data exhibit a significant performance drop when encountering realistic test prefixes. 
To rigorously evaluate fault-revealing capabilities independently of Pass Rates, we performed a comparative analysis on the shared subset of successfully passed problems (Table~\ref{tab:rq1-pitest-qwen} and Table~\ref{tab:rq1-pitest-gpt}). 

The results reveal that achieving higher execution success does not necessarily imply stronger fault-revealing capability. 
For example, although ChatAssert achieves higher Pass Rates than Direct Prompting, its Shared KG is lower on both models. 
This indicates that iterative execution-guided repair may improve executability while potentially weakening assertion strength.

\begin{table}[h!]
\centering
\caption{Evaluation of fault-revealing capability on LLM-based methods. \textbf{Bold} indicates the best result. (a) Results using Qwen3-Coder-30B on 152 shared problems; (b) Results using GPT-5-mini on 183 shared problems.}
\label{tab:rq1-pitest-overall}
\begin{subtable}[t]{0.495\linewidth}
\centering
\caption{Qwen3-Coder-30B}
\label{tab:rq1-pitest-qwen}
\resizebox{\linewidth}{!}{
\setlength{\tabcolsep}{1pt}
\begin{tabular}{lc}
\toprule
Method & Shared KG (\%) \\
\midrule
Trivial & 39.83 \\
Direct Prompting & 92.77 \\
ChatAssert \cite{chatassert} & 87.54 \\
DCAware & \textbf{93.72} \\
\bottomrule
\end{tabular}
}
\end{subtable}
\hfill 
\begin{subtable}[t]{0.495\linewidth}
\centering
\caption{GPT-5-mini}
\label{tab:rq1-pitest-gpt}
\resizebox{\linewidth}{!}{
\setlength{\tabcolsep}{1pt}
\begin{tabular}{lc}
\toprule
Method & Shared KG (\%) \\
\midrule
Trivial & 41.88 \\
Direct Prompting & 91.65 \\
ChatAssert \cite{chatassert} & 86.31 \\
DCAware & \textbf{92.83} \\
\bottomrule
\end{tabular}
}
\end{subtable}
\end{table}

A closer analysis of ChatAssert provides further evidence for the ``Self-Repair Trap''. 
While iterative repair improves execution success, the oracles that first pass at later stages tend to exhibit weaker fault-revealing capability. 

\begin{figure}[h!]
\centering
\includegraphics[width=0.99\linewidth]{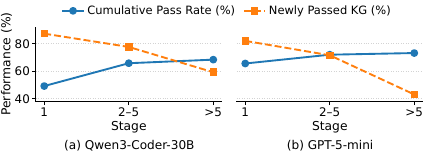}
\caption{Evolution of ChatAssert across generation and repair stages. Stage 1 denotes the initial generation. Pass Rate is cumulative, whereas KG is calculated over the oracles that first pass within each stage group.}
\Description{Two line charts showing ChatAssert results for Qwen3-Coder-30B and GPT-5-mini across three stage groups. Stage 1 represents initial generation, and later stages represent iterative repair. For both models, cumulative Pass Rate increases across stages, while the KG of newly passed oracles decreases.}
\label{fig:chatassert-stage}
\end{figure}

As shown in Figure~\ref{fig:chatassert-stage}, later stages continue increasing cumulative Pass Rate but recover increasingly weaker oracles. 
This suggests that execution feedback can bias iterative repair toward easier-to-satisfy assertions.

In contrast, DCAware avoids iterative repair by constructing high-signal contextual information before generation. 
It achieves the highest overall Pass Rate while maintaining the strongest Shared KG among evaluated LLM-based methods ($93.72\%$ for Qwen3-Coder-30B and $92.83\%$ for GPT-5-mini). 
These results suggest that improving contextual grounding can alleviate the trade-off between execution success and fault-revealing capability.

\begin{findingbox}
\textbf{Answer RQ1:} \textbf{DCAware} outperforms all evaluated baselines in both execution success and fault-revealing capability. 
The stage-wise analysis further shows that iterative repair approaches such as \textbf{ChatAssert} may improve Pass Rates while recovering weaker oracles at later stages. 
DCAware avoids this trade-off through a non-iterative, context-aware generation pipeline. 
\end{findingbox}

\subsection{RQ2: Contribution of Individual Static Contextual Components}
\label{subsec:rq2-static}

\begin{table*}[ht!]
\caption{Execution-based performance of the static-context ablations with dynamic context disabled. Full retains the test class, focal class, and callee components.}
\label{tab:rq2-main}
\setlength{\tabcolsep}{12pt}
\begin{tabular}{llccccc}
\toprule
Model & Method & Pass Rate (\%) & KG (\%) & RE (\%) & TF (\%) & EO (\%) \\
\midrule
\multirow{4}{*}{Qwen3-Coder-30B}
& w/o Test Class    & 68.00             & 59.36             & 12.80 & 19.20 & 0.00 \\
& w/o Focal Class   & 65.60             & 58.57             & 11.40 & 23.00 & 0.00 \\
& w/o Callees       & 70.20             & 62.00             & 7.00  & 22.80 & 0.00 \\
& Full              & \textbf{72.40}    & \textbf{65.92}    & 6.00  & 21.60 & 0.00 \\
\midrule
\multirow{4}{*}{GPT-5-mini}
& w/o Test Class    & 76.60             & 66.18             & 9.00  & 14.20 & 0.20 \\
& w/o Focal Class   & 74.20             & 65.41             & 13.00 & 12.80 & 0.00 \\
& w/o Callees       & 78.00             & 69.22             & 6.80  & 15.00 & 0.20 \\
& Full              & \textbf{80.40}    & \textbf{70.23}    & 5.80  & 13.60 & 0.20 \\
\bottomrule
\end{tabular}
\end{table*}

In this section, we conduct a granular analysis to investigate the specific role of each component within the static context. 
To ensure computational efficiency and minimize interference from dynamic factors, all experiments in this subsection are conducted by excluding dynamic context. 
We compare the following four configurations:
\begin{itemize}
\item \textbf{w/o Test Class:} Removes the skeleton information of the test class, providing only the source code of the test prefix.

\item \textbf{w/o Focal Class:} Excludes the focal class skeleton, retaining only the source code of the focal method.

\item \textbf{w/o Callees:} Strips away other invocations within the test prefix, with the exception of the focal method.

\item \textbf{Full}: Represents the complete static context configuration, which corresponds to the \textbf{w/o Dynamic Context} variant discussed in Section~\ref{subsec:ablation}.
\end{itemize}
The detailed results are demonstrated in Table~\ref{tab:rq2-main}.

The experimental results reveal that the inclusion of test class skeletons and focal class skeletons significantly reduces the RE rate. 
This suggests that these components provide critical references for dependency attributes and method signatures, thereby preventing the model from invoking non-existent properties or methods. 
In contrast, while the contribution of callees is relatively marginal, it still provides auxiliary context that enhances overall performance.

In accordance with Section~\ref{subsection:rq1}, we conduct a mutation analysis of the four configurations.
The detailed results are shown in Table~\ref{tab:rq2-pitest-qwen} and Table~\ref{tab:rq2-pitest-gpt}.

\begin{table}[htbp]
\centering
\caption{Evaluation of fault-revealing capability of four static context configurations. (a) Results using Qwen3-Coder-30B on 176 shared problems; (b) Results using GPT-5-mini on 213 shared problems.}
\label{tab:rq2-pitest-overall}
\begin{subtable}[t]{0.495\linewidth}
\centering
\caption{Qwen3-Coder-30B}
\label{tab:rq2-pitest-qwen}
\setlength{\tabcolsep}{1pt}
\begin{tabular}{lc}
\toprule
Method & Shared KG (\%) \\
\midrule
Trivial & 36.73 \\
w/o Test Class & 92.56 \\
w/o Focal Class & \textbf{93.34} \\
w/o Callees & 91.72 \\
Full & 92.83 \\
\bottomrule
\end{tabular}
\end{subtable}
\hfill
\begin{subtable}[t]{0.495\linewidth}
\centering
\caption{GPT-5-mini}
\label{tab:rq2-pitest-gpt}
\setlength{\tabcolsep}{1pt}
\begin{tabular}{lc}
\toprule
Method & Shared KG (\%) \\
\midrule
Trivial & 38.10 \\
w/o Test Class & 90.25 \\
w/o Focal Class & 91.05 \\
w/o Callees & \textbf{92.28} \\
Full & 92.01 \\
\bottomrule
\end{tabular}
\end{subtable}
\end{table}

The results reveal an inherent trade-off between context size and fault-revealing intensity across both models. While the \textbf{Full} configuration achieves the highest Pass Rates, it ranks second in Shared KG (92.83\% for Qwen3-Coder-30B and 92.01\% for GPT-5-mini). Specifically, the \textbf{w/o Focal Class} setting peaks at 93.34\% for Qwen3-Coder-30B, whereas \textbf{w/o Callees} reaches 92.28\% for GPT-5-mini. 

This indicates that expanding context to resolve dependencies successfully maximizes execution success but introduces minor attentional noise. This interference slightly dilutes the models' focus on generating the most rigorous semantic checks, preventing the \textbf{Full} setting from achieving the absolute highest mutation scores. Overall, this demonstrates the utility of static components in balancing compilation correctness against checking capability, highlighting the inherent trade-offs in context expansion.

To some extent, our findings challenge the conclusions presented in the previous work \cite{molinelli2025llms}. 
This discrepancy can be attributed to two key factors:
\begin{itemize}
\item \textbf{Context Optimization:} Our context compression strategy (i.e., folding uninvoked method bodies) effectively prevents context overflow and provides the model with more precise and relevant information.

\item \textbf{Evaluation Metrics:} Unlike the restrictive EM used in prior work, our execution-based metrics offer a more authentic and accurate assessment of oracle quality, leading to a more reliable measurement of their fault-revealing potential.
\end{itemize}

\begin{findingbox}
\textbf{Answer RQ2:} Integrating structural skeletons balances execution correctness and checking intensity. 
While the \textbf{Full} context maximizes Pass Rates across both models, its expanded size introduces a trade-off: minor attentional noise slightly dilutes assertion strictness, yielding the second-highest fault-revealing performance.
\end{findingbox}

\subsection{RQ3: Contribution of Different Dynamic Contextual Strategies}

\begin{table*}[ht!]
\caption{Execution-based performance of dynamic-context strategies built on the full static context. None provides no dynamic state, Locals dumps available local-variable values, and DCAware retrieves LLM-selected expressions.}
\label{tab:rq3-main}
\setlength{\tabcolsep}{12pt}
\begin{tabular}{llccccc}
\toprule
Model & Method & Pass Rate (\%) & KG (\%) & RE (\%) & TF (\%) & EO (\%) \\
\midrule
\multirow{3}{*}{Qwen3-Coder-30B}
& None & 72.40 & 65.92 & 6.00 & 21.60 & 0.00 \\
& Locals & 72.40 & 66.05 & 5.40 & 22.00 & 0.20 \\
& DCAware & \textbf{80.80} & \textbf{73.20} & 7.40 & 11.80 & 0.00 \\
\midrule
\multirow{3}{*}{GPT-5-mini}
& None & 80.40 & 70.23 & 5.80 & 13.60 & 0.20 \\
& Locals & 84.40 & 71.64 & 4.80 & 10.60 & 0.20 \\
& DCAware & \textbf{88.20} & \textbf{78.17} & 4.20 & 7.60  & 0.00 \\
\bottomrule
\end{tabular}
\end{table*}

In this section, we investigate the contribution of dynamic context in enhancing oracle generation.
Building upon the \textbf{Full Static Context} configuration, we conduct a comparative analysis of the following three strategies:
\begin{itemize}
\item \textbf{None:} No dynamic context is provided. 
Note that this configuration is methodologically identical to the \textbf{Full} static configuration evaluated in RQ2 and the \textbf{w/o Dynamic Context} variant discussed in RQ5.

\item \textbf{Locals:} A naive approach where all available function parameters and local variable values are directly concatenated into the prompt context.

\item \textbf{DCAware:} Uses the full static context together with
the Dynamic Context-Aware phase. 
In this configuration, the LLM autonomously identifies and retrieves the values of specific variables, expressions, or object states it deems necessary during the runtime execution. 
\end{itemize}
The detailed results are shown in Table~\ref{tab:rq3-main}.

Our findings indicate that \textbf{DCAware} significantly contributes to the improvement of Pass Rates. 
Conversely, the naive provision of all local variables (\textbf{Locals}) does not consistently yield performance gains. 
For instance, compared to the baseline with no dynamic context, the Locals strategy shows no improvement in the Pass Rate on Qwen3-Coder-30B and only a marginal increase on GPT-5-mini.

\begin{figure}[h!]
\centering
\includegraphics[width=0.99\linewidth]{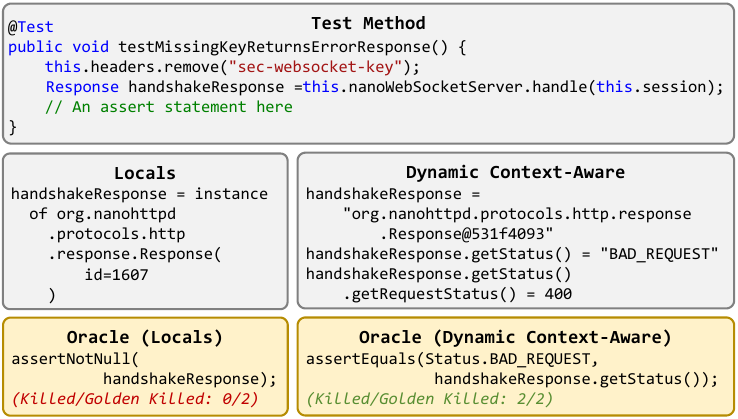}
\caption{Case study comparing Locals and DCAware. Targeted runtime queries enable DCAware to generate a stronger oracle than a raw local-variable dump.}
\Description{Case-study comparison of the Locals strategy and DCAware. Targeted runtime queries expose semantically meaningful object states and enable a stronger fault-revealing oracle than the raw local-variable dump.}
\label{fig:example1}
\end{figure}

Through a detailed case study shown in Figure~\ref{fig:example1}, we observed that \textbf{Locals} often inundates the LLM with raw data that lacks semantic relevance to the specific assertion logic. 
Furthermore, \textbf{Locals} fails to provide access to critical runtime information such as method return values or object attributes at a given execution state. 
In contrast, \textbf{DCAware} enables the LLM to selectively retrieve the values of specific variables and complex expressions it deems critical, thereby providing more targeted and actionable information for generating correct oracles.

\begin{table}[h!]
\centering
\caption{Evaluation of fault-revealing capability of different dynamic context strategies. (a) Results using Qwen3-Coder-30B on 223 shared problems; (b) Results using GPT-5-mini on 263 shared problems.}
\label{tab:rq3-pitest-overall}
\begin{subtable}[t]{0.495\linewidth}
\centering
\caption{Qwen3-Coder-30B}
\label{tab:rq3-pitest-qwen}
\begin{tabular}{lc}
\toprule
Method & Shared KG (\%) \\
\midrule
Trivial & 35.68 \\
None & 92.66 \\
Locals & 93.13 \\
DCAware & \textbf{95.58} \\
\bottomrule
\end{tabular}
\end{subtable}
\hfill
\begin{subtable}[t]{0.495\linewidth}
\centering
\caption{GPT-5-mini}
\label{tab:rq3-pitest-gpt}
\begin{tabular}{lc}
\toprule
Method & Shared KG (\%) \\
\midrule
Trivial & 37.35 \\
None & 90.87 \\
Locals & 88.66 \\
DCAware & \textbf{92.32} \\
\bottomrule
\end{tabular}
\end{subtable}
\end{table}

Similarly, we conduct a mutation analysis on the common subset of problems successfully passed by all evaluated dynamic strategies. 
The results are illustrated in Table~\ref{tab:rq3-pitest-qwen} and Table~\ref{tab:rq3-pitest-gpt}, respectively.

We observe a critical phenomenon validating our Intent-Driven extraction. For GPT-5-mini, providing a large number of raw local-variable values (\textbf{Locals}) improves the Pass Rate but paradoxically causes a sharp decline in fault-revealing intensity compared to the static-only baseline. 
Because \textbf{Locals} inundates the LLM with raw, unfiltered data that lacks semantic relevance, the noise obstructs the model's semantic focus. Furthermore, it fails to capture critical method return states. 
In contrast, \textbf{DCAware} allows the LLM to act as an active debugger, retrieving only targeted, high-entropy expressions, achieving simultaneous improvements in both metrics.

\begin{findingbox}
\textbf{Answer RQ3:} The \textbf{Intent-Driven Dynamic State Extraction} strategy significantly outperforms the naive \textbf{Locals} approach. 
Dumping all available local-variable values introduces noise that degrades fault-revealing intensity (especially in GPT-5-mini), whereas targeted, active querying empowers the LLM to formulate precise semantic assertions.
\end{findingbox}

\subsection{RQ4: Overhead of DCAware}

Token and time consumption are crucial considerations within the context of LLM-based software engineering.
In this section, we calculate the average token usage and time expenditure per problem for all evaluated LLM-based methods.
The detailed results are shown in Table~\ref{tab:rq4-main}.

\begin{table*}[h!]
\caption{Average per-instance token usage and execution time for the LLM-based methods. Prompt, Cached, and Completion are average token counts.}
\label{tab:rq4-main}
\setlength{\tabcolsep}{12pt}
\begin{tabular}{llcccc}
\toprule
Model & Method & Prompt & Cached & Completion & Seconds \\
\midrule
\multirow{3}{*}{Qwen3-Coder-30B}
& Direct Prompting & 349.14 & 18.21 & 19.15 & 0.20  \\
& ChatAssert \cite{chatassert}  & 8881.96 & 3156.73 & 1474.63 & 18.08 \\
& DCAware                       & 6796.23   & 3369.95   & 148.16    & 5.73  \\
\midrule
\multirow{3}{*}{GPT-5-mini}
& Direct Prompting              & 349.10    & 0.00      & 30.75     & 4.21  \\
& ChatAssert \cite{chatassert}  & 10271.60  & 840.03    & 1941.51   & 51.06 \\
& DCAware                       & 6816.10   & 1296.90   & 163.74    & 11.78 \\
\bottomrule
\end{tabular}
\end{table*}

Since Qwen3-Coder-30B is deployed locally, we report the monetary cost of GPT-5-mini.
The monetary costs are calculated based on the standard pricing rates of the Azure API for the GPT-5-mini model. Since the reported total prompt tokens include the cached tokens, the exact cost per request is computed as follows: the uncached prompt tokens (Total Prompt minus Cached) are billed at 0.25 USD per 1M, the cached prompt tokens at 0.025 USD per 1M, and the completion tokens at 2.00 USD per 1M. 
Given that the average cost per problem is extremely small in magnitude, we report the total cost aggregated over the 500 problems in Table~\ref{tab:rq4-cost}.

\begin{table}[h!]
\caption{Total GPT-5-mini API cost over all 500 assertion-completion instances under the official pricing.}
\label{tab:rq4-cost}
\begin{tabular}{lc}
\toprule
Method & Total Cost (\$) \\
\midrule
Direct Prompting & 0.07 \\
ChatAssert \cite{chatassert} & 3.13 \\
DCAware & 0.87 \\
\bottomrule
\end{tabular}
\end{table}

While \textbf{DCAware} constructs a comprehensive structural prompt, its non-iterative, feed-forward design drastically curtails the generation of output tokens compared to \textbf{ChatAssert} (e.g., 163.74 vs. 1941.51 completion tokens for GPT-5-mini). In modern LLM inference, the generation phase (decoding) is significantly more latency-bound and computationally expensive than prompt processing (pre-filling). 
By shifting the computational burden from iterative output generation to high-SNR input construction, \textbf{DCAware} achieves superior generation quality with substantially lower overall execution time and monetary cost.
Notably, \textbf{DCAware} achieves higher prompt caching than the iterative \textbf{ChatAssert}. We achieve this by anchoring the token-heavy static context at the prompt's beginning, ensuring this prefix remains stable across phases to maximize KV-cache hits and minimize costs.

\begin{findingbox}
\textbf{Answer RQ4:} By substituting iterative generation loops with a high-signal non-iterative pipeline, \textbf{DCAware} shifts the computational paradigm. It achieves superior generation quality with approximately one quarter of the API generation cost and up to 77\% less execution time compared to complex agentic frameworks like \textbf{ChatAssert}.
\end{findingbox}

\subsection{RQ5: Ablation Study}
\label{subsec:ablation}
While RQ2 and RQ3 compare alternative designs within the static and dynamic modules, respectively, RQ5 removes complete modules to examine their dependency and complementarity. 
Specifically, we investigate whether dynamic querying relies on static anchors and whether combining both contexts improves the complete \textbf{DCAware} pipeline. 
For compact presentation, we denote the variants that remove the static and dynamic modules as \textbf{w/o Static} and \textbf{w/o Dynamic}, respectively. We evaluate the following configurations: 
\begin{itemize}
\item \textbf{w/o Static}: Omits the Static Context-Aware phase. Similar to \textbf{Direct Prompting}, only the focal method, test setup, and test prefix are provided, while the Dynamic Context-Aware phase remains enabled. 
\item \textbf{w/o Dynamic}: Omits the Dynamic Context-Aware phase. The complete static context is retained, but no dynamic information is provided. This configuration is structurally identical to the \textbf{Full} configuration in RQ2 and the \textbf{None} configuration in RQ3. 
\item \textbf{DCAware}: Uses both the Static Context-Aware and Dynamic Context-Aware phases, representing the complete approach. 
\end{itemize}
The detailed results are shown in Table~\ref{tab:rq5-main}.

\begin{table*}[h!]
\caption{Execution-based comparison of the full DCAware pipeline with complete-module ablations. w/o Static and w/o Dynamic remove the corresponding context-aware phase.}
\label{tab:rq5-main}
\setlength{\tabcolsep}{12pt}
\begin{tabular}{llccccc}
\toprule
Model & Method & Pass Rate (\%) & KG (\%) & RE (\%) & TF (\%) & EO (\%) \\
\midrule
\multirow{3}{*}{Qwen3-Coder-30B}
& w/o Static & 64.00 & 59.81 & 21.60 & 14.40 & 0.00  \\
& w/o Dynamic & 72.40 & 65.92 & 6.00 & 21.60 & 0.00 \\
& DCAware & \textbf{80.80} & \textbf{73.20} & 7.40  & 11.80 & 0.00 \\
\midrule
\multirow{3}{*}{GPT-5-mini}
& w/o Static & 77.40 & 68.82 & 18.20 & 4.40  & 0.00  \\
& w/o Dynamic & 80.40 & 70.23 & 5.80  & 13.60 & 0.20  \\
& DCAware & \textbf{88.20} & \textbf{78.17} & 4.20 & 7.60 & 0.00 \\
\bottomrule
\end{tabular}
\end{table*}

Our ablation study reveals two key findings. 
First, both the static and dynamic contexts contribute significantly to the overall performance, demonstrating the efficacy of both modules. 
Second, removing the static context yields the most substantial impact. This is partially because the static context provides the LLM with more accurate awareness targets. 
In practice, if only the Dynamic Context-Aware phase is enabled without the static context, the model often struggles to identify the most appropriate objects for observation.
This confirms that the static context provides the necessary anchor points; without this architectural blueprint, the dynamic query mechanism operates blindly.

In accordance with Section~\ref{subsection:rq1}, we conduct a mutation analysis. 
The detailed results are shown in Table~\ref{tab:rq5-pitest-qwen} and Table~\ref{tab:rq5-pitest-gpt}.

\begin{table}[h!]
\centering
\caption{Evaluation of the fault-revealing capability of ablated variants. (a) Results using Qwen3-Coder-30B on 178 shared problems; (b) Results using GPT-5-mini on 215 shared problems.}
\label{tab:rq5-pitest-overall}
\begin{subtable}[t]{0.495\linewidth}
\centering
\caption{Qwen3-Coder-30B}
\label{tab:rq5-pitest-qwen}
\setlength{\tabcolsep}{2pt}
\begin{tabular}{lc}
\toprule
Method & Shared KG (\%) \\
\midrule
Trivial & 40.92 \\
w/o Static & 93.90 \\
w/o Dynamic & 92.78 \\
DCAware & \textbf{96.15} \\
\bottomrule
\end{tabular}
\end{subtable}
\hfill
\begin{subtable}[t]{0.495\linewidth}
\centering
\caption{GPT-5-mini}
\label{tab:rq5-pitest-gpt}
\setlength{\tabcolsep}{2pt}
\begin{tabular}{lc}
\toprule
Method & Shared KG (\%) \\
\midrule
Trivial & 41.31 \\
w/o Static & 91.83 \\
w/o Dynamic & \textbf{92.08} \\
DCAware & 92.07 \\
\bottomrule
\end{tabular}
\end{subtable}
\end{table}

On the commonly passed problems, the ablation results further demonstrate the complementary roles of static and dynamic contexts.
For Qwen3-Coder-30B, combining both contexts achieves the highest Shared KG.
For GPT-5-mini, DCAware and the \textbf{w/o Dynamic Context} variant achieve essentially identical Shared KG values (92.07\% vs.\ 92.08\%); therefore, we do not interpret the 0.01-point difference as meaningful.
Nevertheless, dynamic context substantially improves the Pass Rate from 80.40\% to 88.20\% without a material reduction in fault-revealing capability.

\begin{findingbox}
\textbf{Answer RQ5:} Static context provides important anchors for intent-driven dynamic querying, while dynamic context improves the overall execution success of the complete pipeline.
Combining both contexts achieves the strongest overall performance, maximizing Shared KG for Qwen3-Coder-30B and substantially improving the Pass Rate of GPT-5-mini while maintaining comparable Shared KG.
\end{findingbox}

\section{Discussion}

\subsection{The Reward Hacking Phenomenon in LLMs}
Our RQ1 results reveal reward-hacking-like behavior in the evaluated execution-guided regression-oracle workflow. 
When an LLM is placed within a ``Generate-Execute-Repair'' loop, its primary optimization objective implicitly shifts from ``generating a rigorous semantic check'' to ``generating code that compiles and passes''. Consequently, when faced with a complex assertion failure, rather than deeply reasoning about the faulty logic, the LLM takes the path of least resistance. It drops dimensional checks, strips away object field validations, and relies on trivial state verifications (e.g., \verb|assertNotNull| instead of \verb|assertEquals|). 
These results suggest that, without careful control, execution feedback can favor passability over assertion strength. 

\subsection{Static vs. Dynamic Synergy}
The ablation study (RQ5) demonstrates that dynamic context alone performs poorly without static support. This emphasizes the fundamental synergy of our approach: the static context is not merely a source of dependency resolution; it is the blueprint for intent. Without the folded code skeleton, the LLM operates blindly and cannot deduce which variables or methods are worth querying during the dynamic phase. The static phase defines the ``what could be checked'', while the dynamic phase confirms ``what is currently true''.

\subsection{Model Tendencies and Trade-offs}
\label{sec:dis-3}
Our evaluation reveals an intriguing divergence in behavior among different LLMs. 
For instance, Qwen3-Coder-30B tends to prioritize the rigor and fault-revealing strength of the oracle, occasionally at the cost of execution Pass Rates. 
Conversely, GPT-5-mini exhibits a strong inclination towards generating executable, high-pass-rate oracles, but sometimes compromises on testing intensity when presented with overwhelming raw dynamic data (e.g., the \textbf{Locals} strategy). 

\begin{table}[h!]
\caption{Evaluation of fault-revealing capability across the two models using DCAware on 273 shared problems.}
\label{tab:dis-1}
\centering
\begin{tabular}{llc}
\toprule
Method & Model & Shared KG (\%) \\
\midrule
Trivial & -                 & 36.47             \\
DCAware & Qwen3-Coder-30B   & \textbf{92.17}    \\
DCAware & GPT-5-mini        & 90.14             \\
\bottomrule
\end{tabular}
\end{table}

To robustly quantify this divergence, we conducted a direct comparison of the fault-revealing capabilities between the two models using the DCAware approach. 
As shown in Table~\ref{tab:dis-1}, by isolating the intersection of problems successfully passed by both models, we obtained a substantial shared set of 273 problems.
The results confirm our observation: Qwen3-Coder-30B achieves a higher Shared KG (92.17\%) compared to GPT-5-mini (90.14\%). 
This quantitatively demonstrates that while GPT-5-mini excels in yielding runnable oracles, it generates slightly weaker assertions compared to Qwen3-Coder-30B. 
These model-dependent tendencies suggest that future frameworks could adapt their prompting strategies to the underlying model; identifying their causes remains future work. 

\subsection{Performance on a Larger Model}
To investigate whether the effectiveness of DCAware extends beyond the models used in the main evaluation, we additionally evaluate it with GPT-5 under the same protocol. 
DCAware achieves a Pass Rate/KG of 92.20/82.17, compared with 61.60/53.46 for Direct Prompting and 73.80/59.61 for ChatAssert. 
These results suggest that the benefit of DCAware is not limited to smaller models, although a more comprehensive evaluation across additional large models remains future work. 

\subsection{The Necessity of Semantic Identification}
A naive approach to utilizing dynamic execution feedback might involve extracting object states and directly asserting them using an \verb|assertEquals|. 
However, our observations indicate that this method is fundamentally flawed. 
Many complex runtime variables return memory addresses or instance identifiers (e.g., \Verb|handshakeResponse="org.nanohttpd.protocols.http.response.Response@531f4093"|). 
Forcing an equals match on these dynamically allocated addresses leads to extremely fragile and trivial oracles that will inevitably fail upon subsequent executions. 
This underscores the necessity of semantic identification: rather than blindly matching raw string outputs, DCAware empowers the LLM to comprehend the underlying object semantics and extract stable, meaningful attributes (such as return values of specific accessor methods) to construct robust and reproducible test oracles.

\section{Threats to Validity}

\subsection{Internal Validity} 
The non-deterministic nature of LLM inference may affect evaluation consistency. 
For models whose inference APIs expose a temperature parameter, we set it to 0.2. The main tables report the first run, and we
independently reran each complete generation-and-evaluation pipeline twice, resulting in three runs per configuration.

\begin{figure}[ht]
\centering
\includegraphics[width=0.99\linewidth]{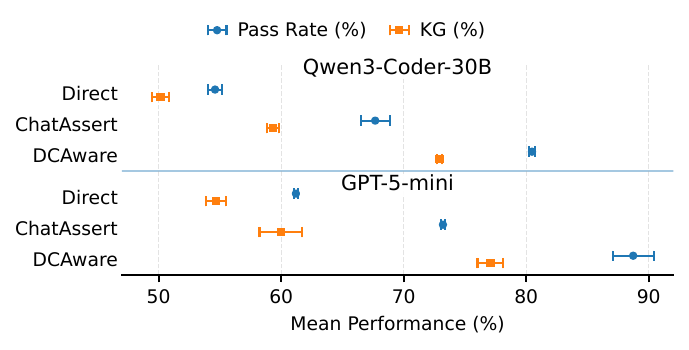}
\caption{Stability of Pass Rate and KG across three independent runs. Markers indicate the mean values, and horizontal error bars denote one standard deviation. The rankings of Direct Prompting, ChatAssert, and DCAware remain unchanged for both evaluated models.}
\Description{A horizontal error-bar plot showing the mean Pass Rate and Kill Golden scores of Direct Prompting, ChatAssert, and DCAware across three independent runs. The upper group reports results for Qwen3-Coder-30B, and the lower group reports results for GPT-5-mini. DCAware achieves the highest mean Pass Rate and Kill Golden score for both models, with relatively small variation across runs.}
\label{fig:stability}
\end{figure}

As shown in Figure~\ref{fig:stability}, the method rankings remain unchanged across the three runs: DCAware consistently achieves the highest Pass Rate and KG, followed by ChatAssert and Direct Prompting. 
The limited standard deviations further indicate that the observed performance differences are stable under repeated generation. 
To assess execution flakiness separately from generation
non-determinism, we additionally executed every golden test and each generated oracle three times and observed no flaky outcomes. All local-model experiments were conducted using the same 40\,GB NVIDIA A100 GPU.

\subsection{Construct Validity}
Construct validity threats primarily concern our measurement methodologies. First, inference engine prefix caching can skew token consumption and latency metrics via unintended cache hits across tasks. We mitigated this by assigning a unique \verb|cache_salt| to each test method's API request, isolating cache memory to ensure accurate resource tracking. 
Prompt guardrails may confound the contribution of dual-context grounding. We therefore applied the same guardrails to Direct Prompting and ChatAssert. 
For Qwen3-Coder-30B, their Pass Rate/KG changed from 55.00/50.76 to 54.40/49.48 and from 68.40/59.00 to 68.20/58.12, respectively; for GPT-5-mini, the corresponding changes were 61.40/54.58 to 59.60/53.39 and 73.20/61.56 to 73.40/56.87. Thus, guardrails alone do not explain DCAware's gains. 

Second, generated test oracles might cover different code paths than the golden oracle, producing an inconsistent set of program mutants. To ensure equitable comparisons, we calculated mutation scores for all automated methods based exclusively on the predefined subset of mutants generated during the human-written golden oracle's execution. 

A potential threat involves the reliability of debugger-based extraction. 
LLM-generated queries may fail due to invalid expressions, unavailable variables, or runtime exceptions. 
In our evaluation, Qwen3-Coder-30B generated an average of 4.75 successful and 2.22 failed queries per instance, while GPT-5-mini generated 7.36 successful and 3.21 failed queries. 
Nevertheless, 88.40\% and 90.60\% of instances obtained at least one successful query for Qwen3-Coder-30B and GPT-5-mini, respectively, indicating that query failures are usually localized rather than preventing dynamic context extraction. 

Regarding side effects, query immutability is enforced through prompt constraints rather than strict static checking. A conservative name-based scan of generated assertions (e.g., \verb|set*|, \verb|remove*|, and \verb|create*|) flagged only 8/500 cases (1.60\%) for each model; since benign calls may match these patterns, this is an upper-bound estimate. Final oracles are evaluated in fresh environments, although JDB attachment may remain challenging for projects with heavy mocking or asynchronous frameworks.

\subsection{External Validity}
A primary threat to external validity is potential data leakage, as the evaluation datasets contain open-source code predating the LLMs' training cutoffs. We addressed this by demonstrating that naive direct prompting yielded much lower accuracy, confirming the models did not perfectly memorize the repositories. 

Furthermore, a significant advantage of our methodology against external validity threats is its structural simplicity. Complex agentic systems (like ChatAssert) often overfit to the strong, multi-turn instruction-following capabilities of specific closed-source models (e.g., GPT-4). Once migrated to smaller or open-source models, these complex prompt chains frequently collapse. 
In contrast, DCAware's fixed, non-iterative pipeline may depend less on model-specific multi-turn behavior, as suggested by its consistent performance on Qwen3-Coder-30B and the evaluated GPT models.

\section{Related Work}
\subsection{Automated Test Oracle Generation}

Automated oracle generation has progressed from information-retrieval and neural methods \cite{atlas,toga,teco} to LLM-based generation \cite{llm-ag,togll}. Subsequent evaluations show that exact-match metrics and synthetic benchmarks can overestimate practical effectiveness, motivating more realistic and execution-based evaluation \cite{hossain2023neural,liu2023towards}. 
Recent empirical and survey work also emphasizes the distinction between intended behavior and regression oracles derived from the current implementation \cite{actual_vs_expected,bodicoat2025understanding,molinelli2025llms, molina2025test}. 
DCAware explicitly targets the latter setting: a test prefix is given, the current program version is treated as expected behavior, and one missing assertion is completed. 

Recent methods improve semantic guidance through fine-tuning \cite{togll}, documentation or natural-language specifications \cite{hossain2025doc2oracll,AugmenTest,endres2024nl2postcond, alagarsamy2025enhancing}, or domain-specific verification \cite{argus}. These approaches primarily target specification alignment or semantic guidance. In contrast, DCAware studies how compressed static dependencies and selectively queried runtime states can strengthen regression-oracle completion without iterative assertion repair. 

\subsection{Execution Feedback in LLM-Based Testing} 
LLMs have been combined with search-based testing and development workflows to generate semantically rich tests \cite{lemieux2023codamosa, sapozhnikov2024testspark}. Program analysis, coverage, mutation, and multi-agent feedback have also been used to guide test generation \cite{altmayer2025coverup,ryan2024code,wang2024hits,nan2025test, dakhel2024effective,candor}, including industrial mutation-testing applications and mutant prioritization \cite{harman2025mutation,bouafif2025primg}. 

Iterative systems such as LLMLOOP, ChatUniTest, and ChatAssert repeatedly generate, execute, and repair code or tests \cite{ravi2025llmloop,chen2024chatunitest,chatassert}. 
ChatAssert is the most closely related approach because it generates assertions using static analysis and execution feedback. DCAware instead uses execution to collect LLM-selected states before oracle generation and does not repair the generated oracle, enabling a fixed pipeline that prioritizes fault revelation and computational efficiency rather than execution success alone. 

\subsection{Contextual Information in Code Generation} 
Repository-level code generation benefits from dependency-aware static context, including local, global, and third-party dependencies \cite{liao2024mathbf,krishna2025codellm}, as well as caller--callee localization \cite{hu2026line}. Dynamic signals have also been incorporated through real-time edit rewards or learned execution feedback \cite{li2024ircoco,sun2024sifting}. 

DCAware combines these directions specifically for regression-oracle completion. Contextual Semantic Folding retains relevant structural dependencies while removing method bodies likely to introduce noise, whereas Intent-Driven Dynamic State Extraction retrieves only expressions selected for the target assertion. Thus, the contribution is not simply adding more context, but coordinating compressed static anchors with targeted runtime observations in a fixed, feed-forward pipeline.

\section{Conclusion}
In this paper, we revisited execution-guided test oracle generation and showed that optimizing for execution success alone is insufficient for producing strong fault-revealing tests. We identified that this objective misalignment can bias the repair process, leading to a feedback-driven degeneration we term the \emph{Self-Repair Trap}. 
Our results suggest that this limitation can be addressed without increasing reasoning complexity. 
Instead, providing high-quality static and dynamic context in a non-iterative pipeline, as instantiated in \textbf{DCAware}, is sufficient to achieve both strong fault detection and efficient generation. 
For the studied regression-oracle completion setting, these findings favor improving contextual quality over adding iterative repair complexity. 
Future work will explore model-specific generation biases for adaptive prompting, address scenarios where the focal method contains latent bugs, and extend intent-driven extraction to more complex repository-level dependencies.

\section*{Data-Availability Statement}
The source code, datasets, and prompts used in this study are available at \cite{li_2026_21357436}.

\bibliographystyle{ACM-Reference-Format}
\bibliography{main}

\end{document}